\documentclass{aa}
\usepackage{graphicx}
\usepackage[round]{natbib}
\usepackage{subfigure}

\begin{document}
\title{Molecular fluorine chemistry in the early Universe}

\titlerunning{Molecular fluorine chemistry in the early Universe}

\author{Denis Puy \inst{1} \and Victor Dubrovich  \inst{1,} \inst{2} \and Anton Lipovka \inst{3}
\and Dahbia Talbi \inst{1} \and Patrick Vonlanthen \inst{1}}

\authorrunning{D. Puy et al.}

\institute{Universit\'e des Sciences Montpellier II, Groupe de Recherche en Astronomie et Astrophysique du Languedoc,
GRAAL CC72, F-34095 Montpellier, France \and SPb Branch of Special Astrophysical Observatory, RAS, St Petersburg, Russia
\and Centro de Investigaci\'on en F\'isica, UNISON, Rosales y Blvd. Transversal, Col. Centro, Edif. 3-I,
Hermosillo, Sonora, M\'exico, 83000
Mexico \\
  \email{Denis.Puy@graal.univ-montp2.fr}
}

\date{Received / Accepted}

\long\def\Htwo{H$_2$}

\abstract
{ Some models of Big Bang nucleosynthesis suggest that very high baryon density regions
were formed in the early Universe, and generated the production of heavy elements other than
lithium such as fluorine F.}
{We present a comprehensive chemistry of fluorine in the post-recombination epoch.}
{Calculation of F, F$^-$ and HF abundances, as a function of redshift $z$, are carried out.}
{The main result is that the chemical conditions in the early Universe can lead to the
formation of HF. The final abundance of the diatomic molecule HF is predicted to be close to
$3.75 \times 10^{-17}$ when the initial abundance of neutral fluorine F is 10$^{-15}$.}
{ These results indicate that molecules of fluorine HF were already present during the
dark age. This could have implications on the evolution of proto-objects and on the anisotropies
of cosmic microwave background radiation. Hydride of fluorine HF may affect enhancement of the emission line intensity from the proto-objects and could produce
spectral-spatial fluctuations. }


   \maketitle

\section{Introduction}
Early Universe chemistry (standard Big Bang chemistry, hereafter SBBC),
of which hydrogen recombination is only one aspect, has been the subject of several studies in the conventional frame of a homogeneous expanding Universe.
One of the most important result of these studies is that the formation of a
significant number of molecules such as $\rm{H_2}$ and HD plays a crucial role on the
dynamical evolution of the first collapsing structures appearing at temperatures
below a few hundred K.\\
\indent
The literature on chemistry in the post-recombination epoch has expanded considerably
in the recent years. Many authors have developed studies of primordial chemistry
in different contexts:
\begin{itemize}
\item Chemical network \cite{lep84}; \cite{puy93}; \cite{sta96};
\cite{gal98}; \cite{lep02}; \cite{pfe03}. \item Formation of the
first objects \cite{puy96}; \cite{abe97}; \cite{abe00};
\cite{shc06}; \cite{nun06}; \cite{hir06}; \cite{yos06};
\cite{puy06}.
\end{itemize}
All these authors clearly point out that trace amounts of light molecules such as $\rm{H_2}$,
HD and LiH were formed during the post-recombination period of the Universe. However
a key item is the question of heavier elements synthesis during the phase of primordial
nucleosynthesis and the formation of the corresponding molecules during the period of
recombination, because many observations suggest that heavy elements already exist in
high redshifts \cite{son96}; \cite{pic03}; \cite{ara04}; \cite{coh04}.\\
\indent It is known that under certain conditions, high density
baryonic bubbles are created in the Affleck-Dine model of
baryogenesis \cite{aff85}, and these bubbles may occupy a
relatively small fraction of space, while the dominant part of the
cosmological volume is characterized by the normal observed
baryon-to-photon ratio. The value of this ratio $\eta$ in the
bubbles could be much larger than the usually accepted value and
still be consistent with the existing data on light element
abundances and the observed angular spectrum of cosmic microwave
background radiation (CMBR). In this context, numerous authors
investigated the possibility that non-homogeneous baryogenesis
produced very high baryon density in a small fraction of the
universe and in these regions some fraction of heavy elements were
already synthesized during primordial nucleosynthesis
\cite{app87}; \cite{tho94}; \cite{jed94}; \cite{kur77};
\cite{jed01}. More recently \cite{mat05} analyzed Big Bang
nucleosynthesis in very high baryon density regions. They found
that primordial synthesis can produce very heavy elements
including proton rich nuclei such as
$^{92}\mathrm{Mo}$, $^{94}\mathrm{Mo}$, $^{96}\mathrm{Ru}$ and $^{98}\mathrm{Ru}$.\\
\indent
Abundance of primary fluorine is below the abundances of the primordial C N O
group during the Big Bang nucleosynthesis. However molecular fluorine could be more
important in comparison with the molecules provided by reactions with C, N and O
atoms. The first detection of interstellar hydrogen fluoride was reported
by \cite{neu97} using the long wavelength spectrometer of ISO satellite. To
interpret this detection of HF, they have constructed a simple model for the steady state of chemistry
of interstellar fluorine with H$_2$ and with water.
Two important factors must be considered for the formation of the molecular component
in the early universe. One is the binding energy of the hydrogen atom in the molecular
component considered, and the second the ionization potential of the molecule. Relative
to these two criteria, the molecule of fluorine HF is an interesting candidate. Ionization
potential of F ($\mathrm{I_{F}}$ = 17.42 eV) is greater than that of hydrogen ($I_{\mathrm{H}}$ = 13.6
eV). Thus
fluorine ions must have recombined before hydrogen ions in the early Universe. Moreover the binding
energy of HF, $\mathrm{E_{HF}}$ = 5.92 eV, is greater than the binding energy of $\mathrm{H_2}$. As
such, HF could be formed in the early Universe.\\
\indent
In this paper we investigate the growth of molecular fluorine abundance. In Sect. 2, we introduce the standard Big Bang
chemistry (hereafter SBBC). In Sect. 3, we perform a detailed analysis of the primordial
chemical evolution of the atomic and molecular species of fluorine. Finally, in Sect. 5 we
discuss some extensions and implications of this study.

\section{Standard Big Bang chemistry}
\label{sect:chem}

The standard Big Bang nucleosynthesis (SBBN) model predicts the nuclei abundances,
mainly those of hydrogen, helium and lithium, and their isotopes in the early Universe.
The standard chemistry of the early Universe is the chemistry of these light elements
and their respective isotopic forms. This chemistry is non-trivial because on one hand
the universal expansion dilutes matter, so lowers the collision rates and slows down the
chemical reactions, while, on the other hand, the matter and radiation temperatures drop,
which decreases the molecule destruction rates, so may encourage molecule formation.\\
\indent After the hydrogen recombination, the ongoing physical
reactions are numerous, partly due to the presence of the cosmic
microwave background radiation (or CMBR), Stancil et al. 1998. We
can establish three classes of chemical reactions:
\begin{itemize}
\item collisional: $\xi + \xi' \longleftrightarrow \xi_1 + \xi_2$
\item electronic: $\xi + \mathrm{e}^- \longleftrightarrow \xi_1 + \xi_2$
\item photo-processes: $\xi + \gamma \longleftrightarrow \xi_1 + \xi_2$.
\end{itemize}

\begin{figure*}[h!tbp]
  \vspace{23mm} \centering
  \includegraphics[width=16cm,height=13cm]{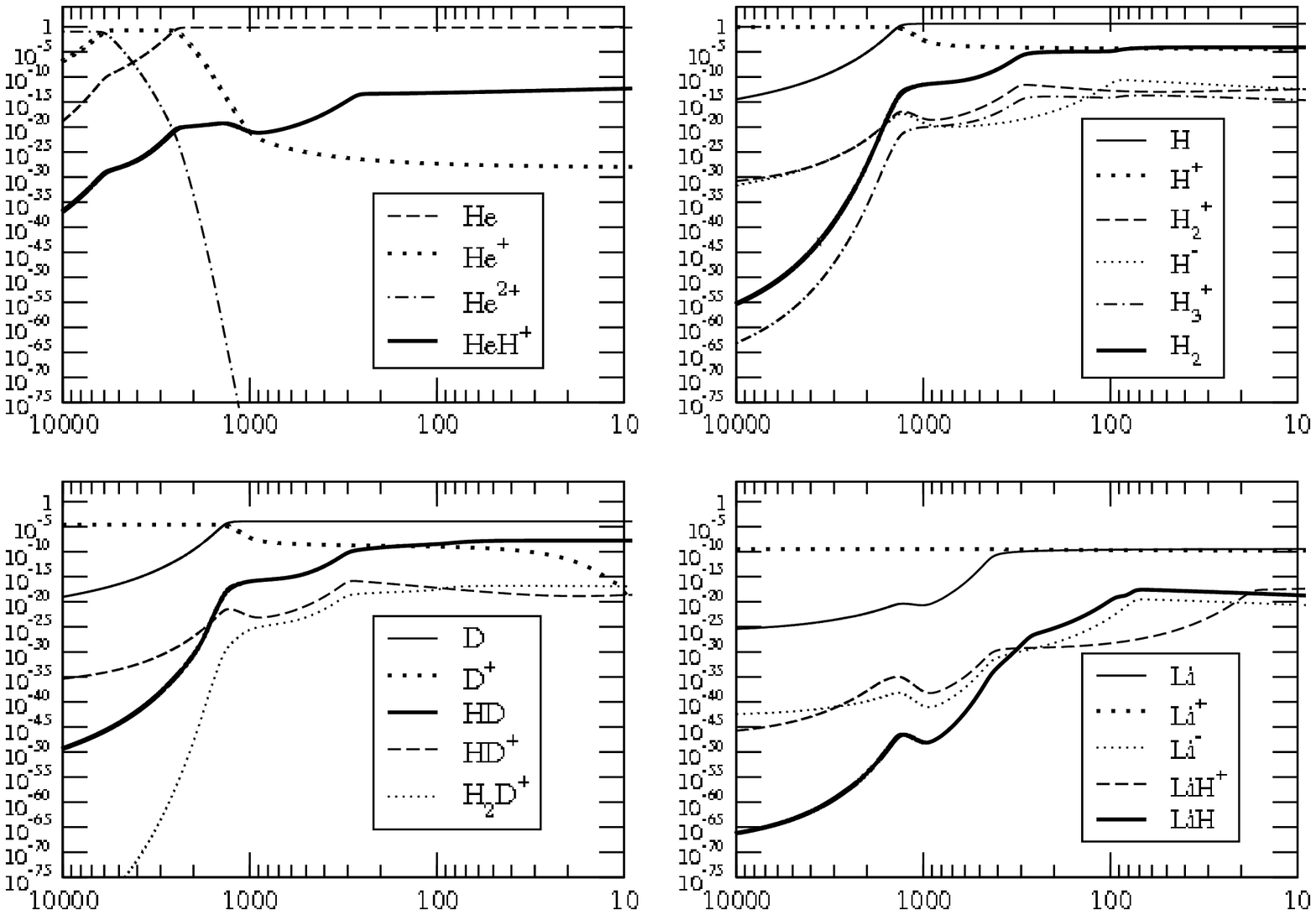}
   \caption{\label{SBBC} Evolution of chemical abundances
in the SBBC model. Vertical axes are the relative abundance and
the horizontal axes are relative to the redshift.}
\end{figure*}

Thus the chemical composition of the primordial gas consists of electrons and:
\begin{itemize}
\item hydrogen: H,H$^-$, H$^+$, H$_{2}^+$ and H$_2$
\item deuterium: D, D$^+$, HD, HD$^+$ and H$_2$D$^+$
\item helium: He, He$^+$, He$^{2+}$ and HeH$^+$
\item lithium: Li, Li$^+$, Li$^-$, LiH and LiH$^+$ .
\end{itemize}
\noindent Their respective abundances are calculated from a set of
chemical reactions for the early Universe. Here we use the
chemical network described by Galli \& Palla (1998). This set of
ordinary differential equations depends on the radiation
temperature $T_{\mathrm{r}}$ or on the matter temperature
$T_{\mathrm{m}}$ and the matter density. All of these evolution
equations must be solved
simultaneously (Puy \& Pfenniger 2006).\\
\indent
The chemical kinetics of a reactant $\xi$ depends on the $\xi$-destruction, by collision with
the reactant $\xi'$ (reaction rate $k_{\xi\xi'}$):
\begin{equation}
\xi + \xi' \rightarrow \xi_1 + \xi_2
\end{equation}
and on the $\xi$-formation process (reaction rate $k_{\xi_1\xi_2}$):
\begin{equation}
\xi_1 + \xi_2 \rightarrow \xi + \xi'
\end{equation}
leading to the kinetic equation of the density $n_{\xi}$:
\begin{equation}
\left( \frac{\mathrm{d}n_{\xi}}{\mathrm{d}t}\right)_{\mathrm{chem}} = \sum_{\xi_1\xi_2} k_{\xi_1\xi_2}
n_{\xi_1} n_{\xi_2} - \sum_{\xi'} k_{\xi\xi'}n_{\xi}n_{\xi'}.
\end{equation}
In the early Universe we must also take into account the decreasing densities due
to the expansion. Thus for each chemical species $\xi$ we have
\begin{equation}
\frac{\mathrm{d}n_{\xi}}{\mathrm{d}t} = -3H(t) \ n_{\xi}+\left(
\frac{\mathrm{d}n_{\xi}}{\mathrm{d}t}\right)_{\mathrm{chem}}.
\end{equation}
The expansion is characterized by the Hubble parameter $H(t)$, which depends on
the energetic component of the Universe:
\begin{equation}
H(t) = H_0 \sqrt{\Omega_{\mathrm{r}} + \Omega_{\mathrm{m}} + \Omega_{\mathrm{K}} +
\Omega_{\mathrm{\Lambda}}}
\end{equation}
where $H_0$=$71 \mathrm{km} \ \mathrm{s}^{-1} \ \mathrm{Mpc}^{-1}$, given by \cite{fre00}, is the present
value of $H(t)$. The $\Omega_i$'s density parameters depend on the redshift $z$ such
as $\Omega_{\mathrm{r}} \equiv (1 + z)^4$ (radiation density parameter) and $\Omega_{\mathrm{m}} \equiv (1
+ z)^3$ (matter density parameter). We consider here a flat Universe (i.e. $\Omega_{\rm{K}}$=0) with a constant
cosmological density parameter $\Omega_{\Lambda} = 0.73$ given by the results of the WMAP experiment
\cite{ben03}.\\
\indent
We extract the initial abundances of atoms and ions out of the SBBN theoretical
predictions given by \cite{cyb03}.
\begin{eqnarray*}
&&[\mathrm{H^+}] \sim 0.889\\
&&[\mathrm{D^+}] \sim 2.436 \cdot 10^{-5}\\
&&[\mathrm{He^{2+}}] \sim 0.111\\
&&[\mathrm{Li^+}] \sim 3.343 \cdot 10^{-10}
\end{eqnarray*}\\
Our calculations start at redshift $z_{\mathrm{i}} = 10^4$, where the considered atoms are fully ionized,
and stop at redshift $z_{\mathrm{f}} = 10$ where the formation of the first structures are
initiated.\\
\indent
In Fig. \ref{SBBC}, we have plotted the chemistry evolution for helium, hydrogen,
deuterium and lithium. Once neutral He is significantly abundant, see Fig. \ref{SBBC}, charge
transfer with ions is possible, allowing the formation of other neutral species, which
permit the formation of $\mathrm{H_2^+}$ through the exchange reaction with some neutral H. Then,
the $\mathrm{H_2^+}$ charge transfer with a neutral species leads to $\mathrm{H_2}$ molecules. Moreover, as
the radiation temperature decreases, $\mathrm{H_2}$ can be formed through $\mathrm{H^-}$ by radiative
attachment followed by associative detachment. Thus the two major routes of $\mathrm{H_2}$ formation are
visible in Fig. \ref{SBBC} by two jumps at the redshifts $z \sim 500$ ($\mathrm{H_2^+}$ channel) and $z \sim
150$ ($\mathrm{H^-}$ channel). HD formation is significant when $\mathrm{H_2}$ appears, the mechanism of
dissociative collision
between $\mathrm{H_2}$ and $\mathrm{D^+}$ ions becomes efficient. LiH is mainly formed by exchange reaction
between Li and $\mathrm{H_2}$, however the charge transfer reactions with the $\mathrm{H^+}$ or H exchange
reaction are important destruction mechanisms of LiH. Thus LiH abundance remains low.

\section{Primordial chemistry of fluorine}
A comprehensive account was given of the chemistry of HF in interstellar clouds by
\cite{zhu02}, and more recently by \cite{neu05} for the chemistry of
fluorine-bearing molecules in diffuse and dense interstellar gas clouds.
One very promising result was presented by \cite{rau94} with a
calculation of heavy element abundances in the context of non-homogeneous Big Bang
nucleosythesis. In this paper the relative abundance of fluorine nuclei is estimated to be close
to $10^{-15}$ which is our initial abundance. Our calculations start at the redshift
$z_{\rm{s}}=2000$ to avoid fluorine recombination. For redshifts greater than $z_{\rm{s}}$, the fluorine
chemistry is essentially dominated by the successive recombination from $\rm{F^{n+}}$ to the neutral
atom F.\\
\indent
We consider seven reactions which involve fluorine component:
\begin{itemize}
\item radiative association: $\rm{F} + \rm{H} \rightarrow \rm{HF} + \gamma$.
\item photodissociation: $\rm{HF} + \gamma \rightarrow \rm{F} + \rm{H}$.
\item radiative attachment: $\rm{F} + \rm{e^-} \rightarrow \rm{F^-} + \gamma$
\item photodetachment: $\rm{F^-} + \gamma \rightarrow \rm{F} + \rm{e^-}$.
\item associative detachment: $\rm{F^-} + \rm{H} \rightarrow \rm{HF} + \rm{e^-}$
\item electronic collision: $\rm{HF} + \rm{e^-} \rightarrow \rm{F^-} + \rm{H}$
\item neutral-neutral reaction: $\rm{F} + \rm{H_2} \rightarrow \rm{HF} + \rm{H}$.
\end{itemize}
In some cases we have estimated the reaction rate of these reactions. All of these reactions
are coupled with the chemical network of the SBBC and the equations of evolution for
the temperatures (matter and radiation) and the density.

\subsection{Radiative association}
The data on the elemental processes of formation-destruction of
molecules are extremely important for the chemical kinetic
calculations. In the case of the HF chemistry one of the possible
channels of the HF formation is radiative association of HF in the
ground electronic state. The only bound electronic state for the
hydrogen halide molecules is the ground $X^1$$\Sigma^+$ state, in
which the molecules can be formed by the radiative association.
There are four states correlating asymptotically with
$\rm{F(^2P)}$+$\rm{H(^2S)}$: the lowest $\rm{X^1 \ \Sigma^+}$
bound state and the three lowest repulsive $^1\Pi$, $^3\Sigma^+$
and $^3\Pi$ states \cite{bro00}. The triplet states are dipole
forbidden to combine radiatively with the singlet bound state,
unless spin-orbit coupling is strong which is not the case
\cite{bro00}. Therefore radiative association from
$\rm{F(^2P)}$+$\rm{H(^2S)}$ is only possible through the bound
$\rm{X^1}$$\rm{\Sigma^+}$ and the repulsive $\rm{A \ ^1\Pi}$
states for the HF formation:
\begin{eqnarray*}
(\mathtt{F1}) \quad \rm{F} + \rm{H} \rightarrow \rm{HF} + \gamma .
\end{eqnarray*}
We can calculate the reaction rate by using the principle of detailed balance in the case
where the cross section of inverse process (photodissociation of the HF) is available.
Brown \& Balint-Kurti (2000) have published data on the total photodissociation
cross section for $A^1\Pi \longleftarrow X^1\Sigma^+$ which can be used to obtain the radiative association
cross section:
\begin{equation}
\sigma_a = \frac{2 g_{\mathrm{n}}}{Z_H Z_F} \left( \frac{h \nu}{\mu c v} \right)^2 \sigma_d,
\end{equation}
Here $g_{\rm{n}}$ is the statistical weight, $Z_H$, $Z_F$ are statistical sums for H and F atoms
respectively, $\nu$ is the frequency of photon, $v$ is the relative velocity of H and F atoms, $\mu$
is the reduced mass of the molecule and $\sigma_d$ is the photodissociation cross section. Finally,
the rate of the radiative association can be obtained by integration over $v$.
\begin{equation}
k_{\mathtt{F1}}(T_{\rm{m}}) = \int\limits^{\infty}_{0} \sigma_a \ f(T_{\rm{m}}, v) \ v \ \mathrm{d}v,
\end{equation}
where $f(T_{\rm{m}}, v)$ is the Maxwell distribution function averaged over all angles and $T_{\rm{m}}$ is
the matter temperature.\\
\indent
We find with high accuracy that the rate $k_{\mathtt{F1}}$ can be fitted by the approximation:
\begin{equation}
k_{\mathtt{F1}} = 1.79 \times 10^{-17} \ T_{\rm{m}}^{-0.51} \ \rm{cm}^3 \ \rm{s}^{-1}.
\end{equation}
\subsection{Photodissociation of HF}
$(\mathtt{F2}) \quad \rm{HF} + \gamma \rightarrow \rm{F} + \rm{H}$
\vskip2mm
\noindent
Using available cross sections from Brown \& Balint-Kurti (2000) for the photodissociation of HF we
calculated the
corresponding rate to be $k_{\mathtt{F2}} = 6.8 \times 10^7 \ \exp \left( -121171/T_{\rm{r}} \right) \ \rm{s^{-1}}$
where $T_{\rm{r}}$ is the radiation temperature.
\subsection{Radiative attachment}
$(\mathtt{F3}) \quad \rm{F} + \rm{e^-} \rightarrow \rm{F^-} +
\gamma$ \vskip2mm \noindent Considering that photodissociation of
F$^-$ should be close to that of Cl$^-$, we deduced the radiative
association cross section of $(\mathtt{F3})$ from the experimental
cross section of $\rm{Cl^-} + \gamma \rightarrow \rm{Cl} +
\rm{e^-}$ \cite{rot69} by using the principle of detail balance.
With this cross section, we calculated the reaction rate, to be
constant, and of the order of $k_{\mathtt{F3}} = 10^{-14} \
\rm{cm}^3 \ \rm{s^{-1}}$.
\subsection{Photodetachment}
$(\mathtt{F4}) \quad \rm{F^-} + \gamma \rightarrow \rm{F} + \rm{e^-}$
\vskip2mm
\noindent
With the same hypothesis as above we used the experimental estimated cross section of the reaction
$\rm{Cl^-} + \gamma \rightarrow \rm{Cl} + \rm{e^-}$ to deduce the rate constant $(\mathtt{F4})$ for the photodetachment.
We find an analytical expression $k_{\mathtt{F4}} = 633120 \ \Bigr(T_{\rm{r}}/300 \Bigl)^{3.6} \ \exp (-41481/T_{\rm{r}}) \ \rm{s^{-1}}$.
\subsection{Associative detachment}
$(\mathtt{F5}) \quad \rm{F^-} + \rm{H} \rightarrow \rm{HF} +
\rm{e^-} \, .$ \vskip2mm \noindent For this reaction we have used
the experimental rate constant of Fehsenfeld \cite{feh73},
determined at 296 K, $k_{\mathtt{F5}} = 1.6 \times 10^{-9}\
\rm{cm}^3 \ \rm{s^{-1}}$.
\subsection{Electronic collision}
$(\mathtt{F6}) \quad \rm{HF} + \rm{e^-} \rightarrow \rm{F^-} + \rm{H}$
\vskip2mm
\noindent
Using the experimental cross section for this process from \cite{xu00}, we calculated the rate constant to be
$k_{\mathtt{F6}} = 8.1 \times 10^{-17} \ \rm{s^{-1}}$ at our temperature of interest.
\subsection{Neutral-neutral reaction}
$(\mathtt{F7}) \quad \rm{F} + \rm{H_2} \rightarrow \rm{HF} + \rm{H}$. The rate coefficient has been measured, see
\cite{umi99}, \textbf{$k_{\mathtt{F7}} = 10^{-10} \ \exp (-400/T_{\rm{m}}) \ \rm{cm^{-3} \ s^{-1}}$.}

\begin{figure*}[h!tbp]
  \vspace{20mm} \centering
\includegraphics[width=16cm,height=12cm]{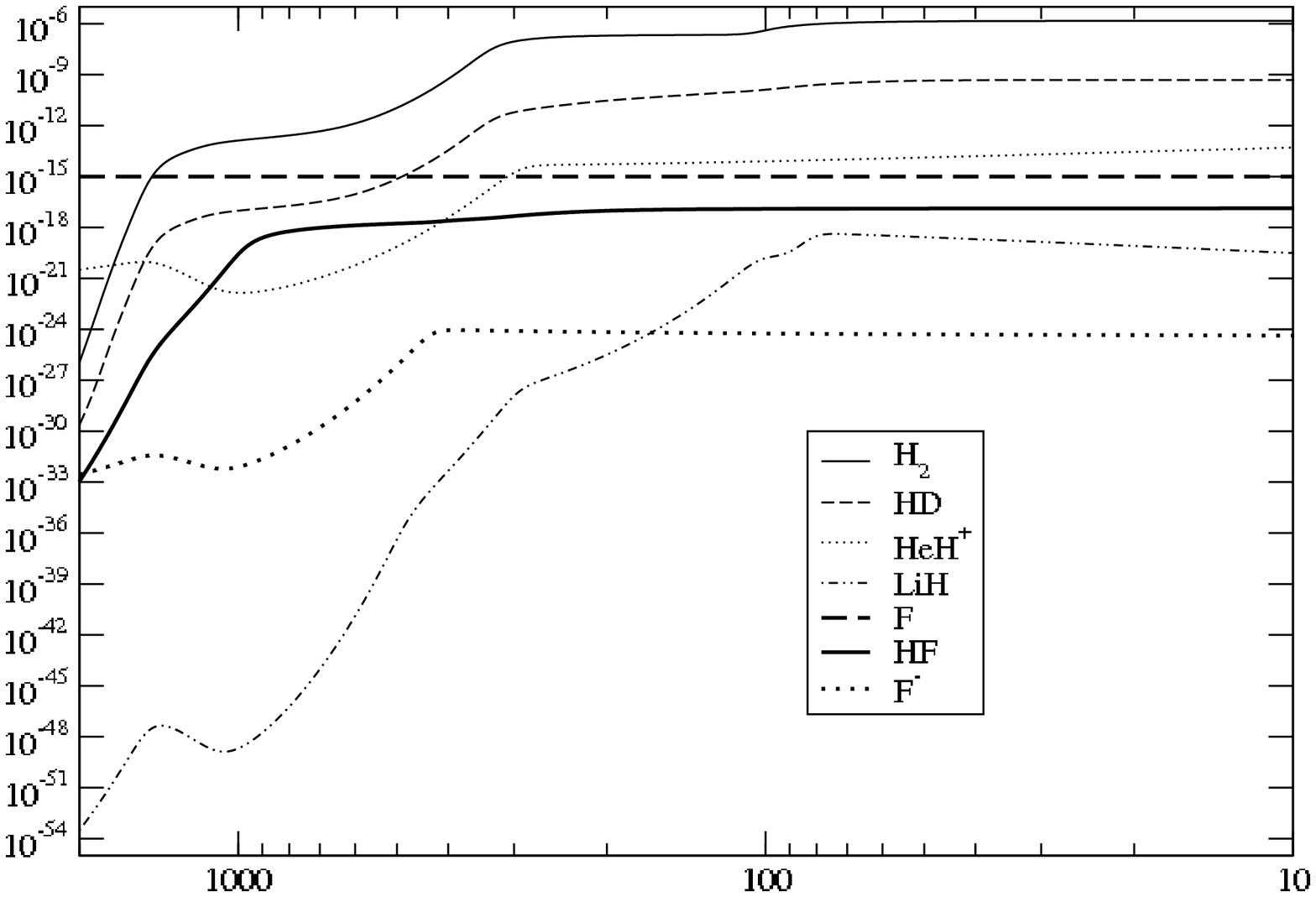}
\caption{\label{HF} Evolution of fluorine abundances compared to
$\rm{H_2}$, HD, LiH and $\rm{HeH^+}$ abundances of SBBC model.
Vertical axis is the relative abundance when the horizontal axis
is relative to the redshift.}
\end{figure*}

\section{Results and discussion}
Figure \ref{HF} shows clearly the formation of HF molecules. The relative abundance of fluorine
species are at $z = 10$:
\textbf{\begin{itemize}
\item{} [F] = $9.90 \times 10^{-16}$
\item{} [HF] = $1.06 \times 10^{-17}$
\item{} [F$^-$] = $4.02 \times 10^{-25}$
\end{itemize}}

HF molecules are mainly formed by the exothermic neutral-neutral reaction of atomic
fluorine with $\mathrm{H_2}$, which is confirmed by \cite{sno06} where they reported results
from a survey of neutral fluorine in the interstellar medium. They showed that the hydride
of fluorine, HF, is in competition with the abundance of atomic fluorine, depending on
how abundant $\mathrm{H_2}$ is.\\
\indent Primordial molecules could play an important role in the
future observations of CMBR anisotropies(\cite{dub77};
\cite{mao94}. These authors argued that huge enhancement of the
emission line intensity from the proto-objects could occur, and
mentioned that molecular protoclouds at high redshift, $10 < z <
300$, which have a peculiar velocity relative to the CMBR, could
produce spectral-spatial fluctuations through the mecanism of pure
reflection of the CMBR photons due to the opacity of a
proto-object in narrow spectral lines and the Doppler shift in
frequency due to the peculiar velocity. In this context, molecules
of fluorine HF could play an interesting role, despite the low
initial abundance of fluorine compared to the abundance of lighter
elements. We find (see Fig. \ref{HF}), that the abundance of
molecular fluorine is the third neutral molecular abundance (with
$\rm{H_2}$ and HD). Thus HF molecules could contribute to the
process of fragmentation of large proto structures. HF
molecules\footnote{where the dipole moment of HF is closed to
$d_{\rm{HF}} = 1.8$ Debye, see \cite{mue72}.} could be an
important coolant agent in collapsing proto-objects and initiate
the mechanism of thermal
instability as do HD and LiH Puy \& Signore 1996, 1997; \cite{lip05}.\\
\indent The detection of primordial HF will be possible in the
near future with instruments such as ALMA or HERSCHEL satellite
\cite{mao05}. Searching for HF could result in constraints on the
fundamental process of nuclei formation during the Big Bang
nucleosynthesis.

\begin{acknowledgements}
  We acknowledge the PNC (\textsf{Programme National de Cosmologie}) and the PNPS (\textsf{Programme
  national de Physique Stellaire}) for their financial assistance, as well as to Corinne Serres-Cousin\'{e} for
  their helpful advices. VD is endebted to the French
  CNRS for financial support through a contract of associated researcher at the University of
  Montpellier II.
\end{acknowledgements}


\end{document}